\documentstyle[{epsf}]{l-aa}

\newcommand{\eal}{et al.\ }
\newcommand{\Lya}{Ly$\alpha$}

\newcommand{\gsim}{\raisebox{-5pt}{$\;\stackrel{\textstyle >}{\sim}\;$}}
\newcommand{\lsim}{\raisebox{-5pt}{$\;\stackrel{\textstyle <}{\sim}\;$}}

\begin{document}
 
   \thesaurus{03         
              (11.17.1);  
              (11.17.4 PKS 2126--158)} 

   \title{The absorption spectrum of the QSO PKS 2126--158 ($z_{\rm em}=3.27$) 
         at high resolution 
   	\thanks{Based on observations collected at the European
Southern Observatory, La Silla, Chile (ESO No. 2-013-49K).} 
	}

   \author{V. D'Odorico\inst{1}
	\and S. Cristiani\inst{2}
	\and S. D'Odorico\inst{3}
	\and A. Fontana\inst{4}
	\and E. Giallongo\inst{4}
          }
 
   \offprints{Stefano Cristiani}
 
   \institute{International School for Advanced Studies, SISSA, via Beirut 
	      2-4, I-34014 Trieste, Italy
	\and Dipartimento di Astronomia dell'Universit\`{a} di Padova,
	      Vicolo dell'Osservatorio 5,\\I-35122 Padova, Italy
	\and European Southern Observatory, Karl Schwarzschild Strasse 2, 
	     D-85748 Garching, Germany
	\and Osservatorio Astronomico di Roma, via dell'Osservatorio, I-00040
	     Monteporzio, Italy
	}
 
   \date{Received \dots; accepted \dots}
 
   \maketitle
 
   \begin{abstract}
%
Spectra of the $z_{em} = 3.268$ quasar PKS 2126--158 have been obtained
in the range $\lambda\lambda 4300-6620$ \AA\ with a resolution $R\simeq27000$ 
and an average signal-to-noise ratio $s/n\simeq 25$ per 
resolution element.
The list of the identified absorption lines is given together with
their fitted column densities and Doppler widths. 
The modal value of the Doppler parameter distribution for the \Lya\ 
lines is $\simeq 25$ km s$^{-1}$. 
The column density distribution can be described by a power-law 
$dn / dN \propto N^{-\beta}$ with $\beta\simeq 1.5$.

12 metal systems have been identified, two of which were previously
unknown. 
In order to make the column densities of the intervening
systems compatible 
with realistic assumptions about the cloud sizes and the silicon to carbon
overabundance, it is necessary to assume a jump beyond the He~II edge
in the spectrum of the UV ionizing background at $z \sim 3$ 
a factor 10 larger than the standard predictions for the
integrated quasar contribution.

An enlarged sample of C~IV absorptions 
(71 doublets) has been used 
to analyze the statistical properties of this class 
of absorbers strictly related to galaxies.
The column density distribution is well described by a 
single power-law, with $\beta=1.64$ and the Doppler parameter 
distribution shows a modal value $b_{\rm CIV} \simeq 14$ km s$^{-1}$. 
The two point correlation function has been computed in the 
velocity space for the individual components of C~IV features. 
A significant signal is obtained for scales smaller than $200-300$ km 
s$^{-1}$, $\xi(30< \Delta v < 90 {\rm km\ s}^{-1}) = 32.71 \pm 2.89$.
A trend of decreasing clustering amplitude with decreasing
column density is apparent, analogously to what has been observed 
for \Lya\ lines.
   \end{abstract}
 
%
 
\section{Introduction}
 
The study of quasars at high redshift has gained, in the last years, a 
remarkable position in cosmology because of its important contribution to 
the knowledge of the formation and evolution of cosmological structures.     
Besides the study of the quasar population 
itself, the analysis of the quasar absorption spectra allows us to probe the 
intervening matter up to the redshift of the emitting object. 

Thanks to the improvement in astronomical instrumentation, a 
considerable amount of high resolution data has become available 
in recent years, greatly increasing the knowledge of the nature and 
evolution of the absorbers. 

The absorption spectra of QSOs at high redshifts are characterized by a
``forest'' of lines shortward of the \Lya\ emission, first noted by Lynds
(1971), who correctly interpreted them as \Lya\ due to absorbers 
distributed along the line of sight. 

The large range in column
densities of these lines suggests the presence of very different intervening
structures, from fluctuations of the diffuse intergalactic medium to the
interstellar medium in protogalactic disks. 

Profile fitting techniques (e.g. Fontana \& Ballester 1995; Hu et al. 1995) 
provide
the few parameters (redshift $z$, Doppler width $b$ and column density $N$)
constituting the basis for the interpretation of the physical properties 
of the absorbers.
The reliability of such parameters depends strongly upon the resolution
and the signal to noise ratio of the spectral data. 
Controversies arisen in the past may be traced back to the difficulty of the
observational problem, at the limit of the present technology
(Pettini \eal 1990; Carswell \eal 1991; Rauch \eal 1993). 

Remarkable results have been recently obtained in the study  of
the clustering properties of \Lya\ clouds.
The absence of clustering 
for velocity separations between 300 and 30\,000 km s$^{-1}$ has been
assessed (Sargent \eal 1980,1982; Bechtold \eal 1987; Webb \& Barcons
1991). 
On the other hand, 
the presence of a significant non-zero correlation function
has been observed for \Lya\ lines with $\Delta v <300$ km s$^{-1}$ and $\log
N_{\rm HI} \ga 13.8$ (Chernomordik 1995; Cristiani \eal 1995; Meiksin \&
Bouchet 1995, Hu et al. 1995; Fern\'andez-Soto et al. 1996). 
The correlation function for the present sample of \Lya\ lines
is discussed in Cristiani \eal (1997). 

Lanzetta \eal (\cite{Lanz95}) inferred from observational results 
that, at $z\le 1$, at least 32 \% (but it could be as high as 60 \%) of the 
\Lya\ absorption systems arise in luminous galaxies. This conclusion 
is at variance with the longstanding belief that \Lya-forest absorption 
systems arise in intergalactic clouds. Moreover, recent spectra at very 
high resolution have revealed the presence of C~IV 
absorptions associated with \Lya\ lines with $\log N_{\rm HI} \ga14$ 
(Cowie \eal 1995; Tytler et al. 1995; Womble et al. 1996; Songaila \& Cowie 
1996).  The derived abundances seem to be similar to the ones derived 
for the heavy elements absorptions originated in galactic halos, suggesting 
continuity in their physical properties. 

The clustering of C~IV absorption systems has been investigated in the 
past using large samples observed at low resolution - FWHM $> 1$ \AA\ -  
(33 QSOs in Young et al. 1982; 55 QSOs in Sargent et al. 1988 ). 
From these data it has been possible to assess that C~IV systems 
do cluster on scales $\Delta v \lsim
600$ km s$^{-1}$, but, due to the limited spectral resolution, the 
characteristic clustering scale and the 
clustering properties at smaller velocity separations have not been 
established.   

Petitjean and Bergeron (1994) analyzed a sample of 10 QSOs with
a higher spectral resolution. They obtained as best fit of the two point 
correlation function in the range 
$30 - 1000$ km s$^{-1}$ a sum of two Gaussian components with 
dispersions of 109 and 525 km s$^{-1}$, similar to the 
distribution observed for Mg~II systems (Petitjean \& Bergeron 1990). 
This result further confirms that metal systems as 
identified by C~IV absorptions features arise in galactic halos. 


More recent works by Womble et al. (1996) and Songaila and 
Cowie (1996) analyzed clustering of lower column 
density samples detecting a lower signal on the same 
scales studied by Petitjean and Bergeron (1994). 
This trend with column density is consistent with what is observed for 
\Lya\ lines (see Cristiani et al. 1997). 


In the present paper we address these issues on the basis of the absorption
spectrum of the quasar PKS 2126--158 ($z_{\rm em}= 3.268$). 
This object is one of the brightest quasars 
known ($m_V \simeq 17.1$) and for this reason its spectrum has been 
studied in many previous 
works (Jauncey \eal 1978; Young \eal 1979; Sargent \& Boksenberg 1983; 
Meyer \& York 1987; Sargent \eal 1988; Sargent \eal 1989; Sargent \eal 1990; 
Giallongo \eal 1993 hereafter GCFT). 
Here we present new data, providing improved quality spectra in terms
of $s/n$ ratio, resolution and wavelength range.

The paper is organized as follows. 
In Sec.~2 we describe the observations and the data reduction, 
in Sec.~3 the method of analysis. 
Section 4 discusses the statistical properties of the \Lya\ sample. 
The description of the metal-line systems is given in Sec.~5. 
Section 6 analyses some statistical properties of the C~IV absorption systems. 
In Sec.~7 the two point correlation function is computed for the C~IV 
lines and the correlation between clustering and column density is 
investigated.
The conclusions of the paper are in Sec. 8.  

\section{Data acquisition and reduction}

In August 1991 and August 1994, 8 echelle observations of the quasar PKS 
2126--158 were obtained at ESO (La Silla), with the NTT telescope and 
the EMMI instrument (see \cite{dodo90}). The ESO echelle \#10 was used in the 
red arm of the instrument in combination with a grism cross-disperser. 
The seeing during the observations 
was typically between 0.8 and 1.3 arcsec.

The absolute flux calibration was carried out by observing the standard 
star EG274 (\cite{sto:bal83}).

The data have been reduced in the context of the ECHELLE software 
package available in the 94NOV edition of MIDAS, the ESO data reduction 
system. Some improvements have been introduced with respect to the standard 
procedure. 
In particular, the spectra obtained from different observations are summed up 
without a previous rebinning but maintaining the original pixel sampling. 
For the wavelength calibration Thorium lamp spectra have been used. 
Wavelengths have then been corrected to vacuum heliocentric. 

The weighted mean of the spectra is equivalent to $\sim 16$ hours of 
integration time and it presents a resolution $R\simeq27000$. 
The $s/n$ ratio per resolution element at the continuum level is 
doubled compared with the previous spectrum by 
GCFT: it ranges from $s/n\simeq 6$ to $44$ in the interval 
$4380-5190$ \AA, while beyond the Ly$\alpha$ emission it has a 
roughly constant value of $s/n\simeq 30$.

In addition to these data, 8 long slit observations of the spectrum of 
PKS 2126--158 were obtained with an holographic grating (ESO \#11) at the blue 
arm of the EMMI instrument. 
The reduction has been carried out with the LONG SLIT package of 
MIDAS, using an optimal extraction option. The resolution of these 
data is $R\sim 8500$. 

The resulting spectrum has been used to check the reliability of the \Lya\  
identification process by verifying the presence of the higher order 
Lyman series lines. 

\begin{figure*}
\caption[1]{\label{f1a} Spectrum of PKS 2126--158 in the wavelength 
range $\lambda\lambda 4300-6620$ covered by the new observations showing 
fits to identified lines overlaid upon the normalized spectrum. 
The spectral interval $\lambda\lambda 6620-7000$ and the corresponding 
absorption lines are shown in GCFT. 
The dashed line represents the noise per resolution element. 
Upper vertical ticks correspond to \Lya~lines, lower 
ticks correspond to  metal lines.}  
\end{figure*}  



The journal of the observations is given in Tab.~\ref{t1}. 

\begin{table}
\caption[t1]{Journal of the observations.}  
{\label{t1}}
\begin{flushleft}
\begin{tabular}{ccccc}
\hline \hline \noalign{\smallskip}
date       & exposure & slit   & range & CCD \\
(yy mm dd) & (s)      & width  & (\AA) &     \\
\noalign{\smallskip}
\hline \noalign{\smallskip}
91 08 06 & 7200 & 1".5 & 4539-7055 & FORD 2k \\
91 08 06 & 7200 & 1".5 & 4539-7055 & FORD 2k \\
91 08 07 & 9130 & 1".5 & 4539-7055 & FORD 2k \\
94 08 13 & 5380 & 1".0 & 4056-6635 & TK 2k \#36 \\
94 08 13 & 7200 & 1".0 & 4056-6635 & TK 2k \#36 \\
94 08 13 & 7200 & 1".0 & 4056-6635 & TK 2k \#36 \\
94 08 15 & 7200 & 1".2 & 4279-6644 & TK 2k \#36 \\
94 08 15 & 6645 & 1".2 & 4279-6644 & TK 2k \#36 \\
94 08 14 & 2700 & 1".2 & 3822-3980 & TK 1k \#31 \\
94 08 14 & 2700 & 1".0 & 3822-3980 & TK 1k \#31 \\
94 08 14 & 2700 & 1".0 & 3686-3846 & TK 1k \#31 \\
94 08 14 & 2700 & 1".0 & 3686-3846 & TK 1k \#31 \\
94 08 15 & 3600 & & 3825-3983 & TK 1k \#31 \\
94 08 15 & 3600 & & 3825-3983 & TK 1k \#31 \\
94 08 15 & 3600 & & 3687-3848 & TK 1k \#31 \\
94 08 15 & 3000 & & 3687-3848 & TK 1k \#31 \\
\noalign{\smallskip}
\hline
\end{tabular}
\end{flushleft}
\end{table} 

\section {The detection and measurement of absorption lines}

\begin{figure*}
\epsfxsize=18cm
\epsffile{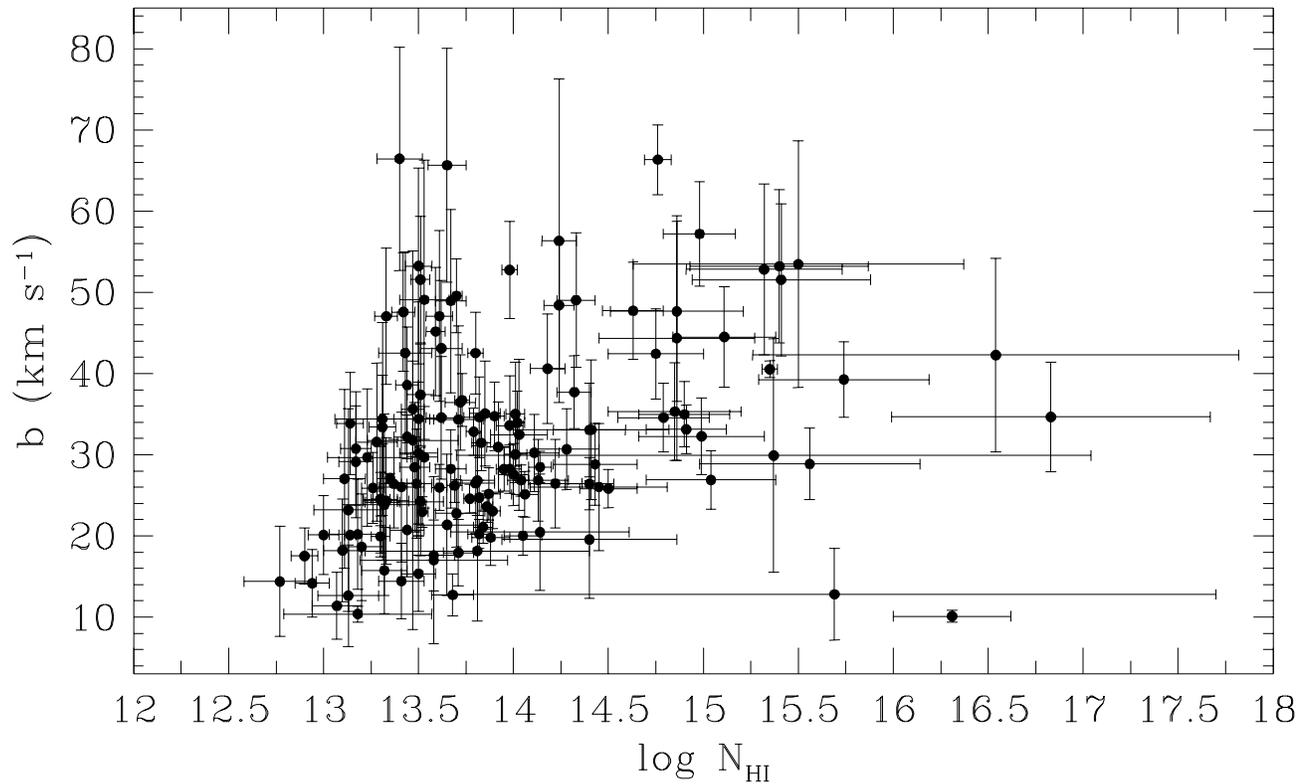}
\vskip -12truecm
\caption[ ]{\label{fig:b-N} Plot of the Doppler parameter $b$ vs. the 
logarithmic column density $\log N_{\rm HI}$ for the Ly$\alpha$ 
lines listed in Tab.~\ref{t2}.} 
\end{figure*}

The determination of the continuum in the QSO spectrum is a critical step 
because it affects the measurement of the absorption line parameters. While 
the continuum redward of \Lya\ emission can be drawn without difficulty, the 
high line density in the Lyman forest complicates the operation
(Young et al. 1979, Carswell et al. 1982).  
In the present work, the task has been fulfilled with the help of a 
procedure, allowing to select automatically and in a 
reproducible way the regions of the spectrum free of strong absorption 
lines or artificial peaks (e.g. due to cosmic rays), i.e. where the RMS 
fluctuation about the mean becomes consistent with noise statistics. 
The continuum level has been estimated by spline-fitting these regions
with quadratic polynomials.
The normalized spectrum is shown in Fig.~\ref{f1a}. 

The detection and measurement of absorption lines in the spectrum have 
been carried out as in GCFT and we refer to this paper for details of the 
procedure. In particular, the lines have been fitted with Voigt profiles 
convolved with the instrumental spread function, making use of a 
minimization method of $\chi^2$. This step has been performed within the 
MIDAS package with the programme FITLYMAN (\cite{fitly95}). The values of 
redshift $z$, Doppler parameter $b=\sqrt{2}\,\sigma$ (where $\sigma$ is 
the velocity dispersion) and column density $N$ 
have been determined for isolated lines and individual components of blends. 

The number of components of each absorption feature is assumed to be the 
minimum required to give a reduced $\chi^2 <1$  ( corresponding to a 
confidence level $P\gsim 50 \%$).

The identification of the metal systems is described in Sect. 5. 

All the lines shortward of the \Lya\ emission not identified as due to metals 
have been fitted as \Lya\ and Ly$\beta$. 
For a further control, we used the 
blue, lower-resolution spectra to search for 
Ly$\beta$, Ly$\gamma$ and Ly$\delta$ lines
in correspondence to the stronger \Lya\ lines ($\log N_{\rm HI}\ge 14$). 
In few cases, we could ascertain that the absorption was not \Lya. 
Such lines are listed as 
unidentified in Tab.~\ref{t2} and they probably belong to metal 
systems still to be recognized.

\begin{table*}\caption[t2]{Absorption line parameters of the \Lya~forest.}
{\label{t2}}
\epsfysize=26cm
\epsffile{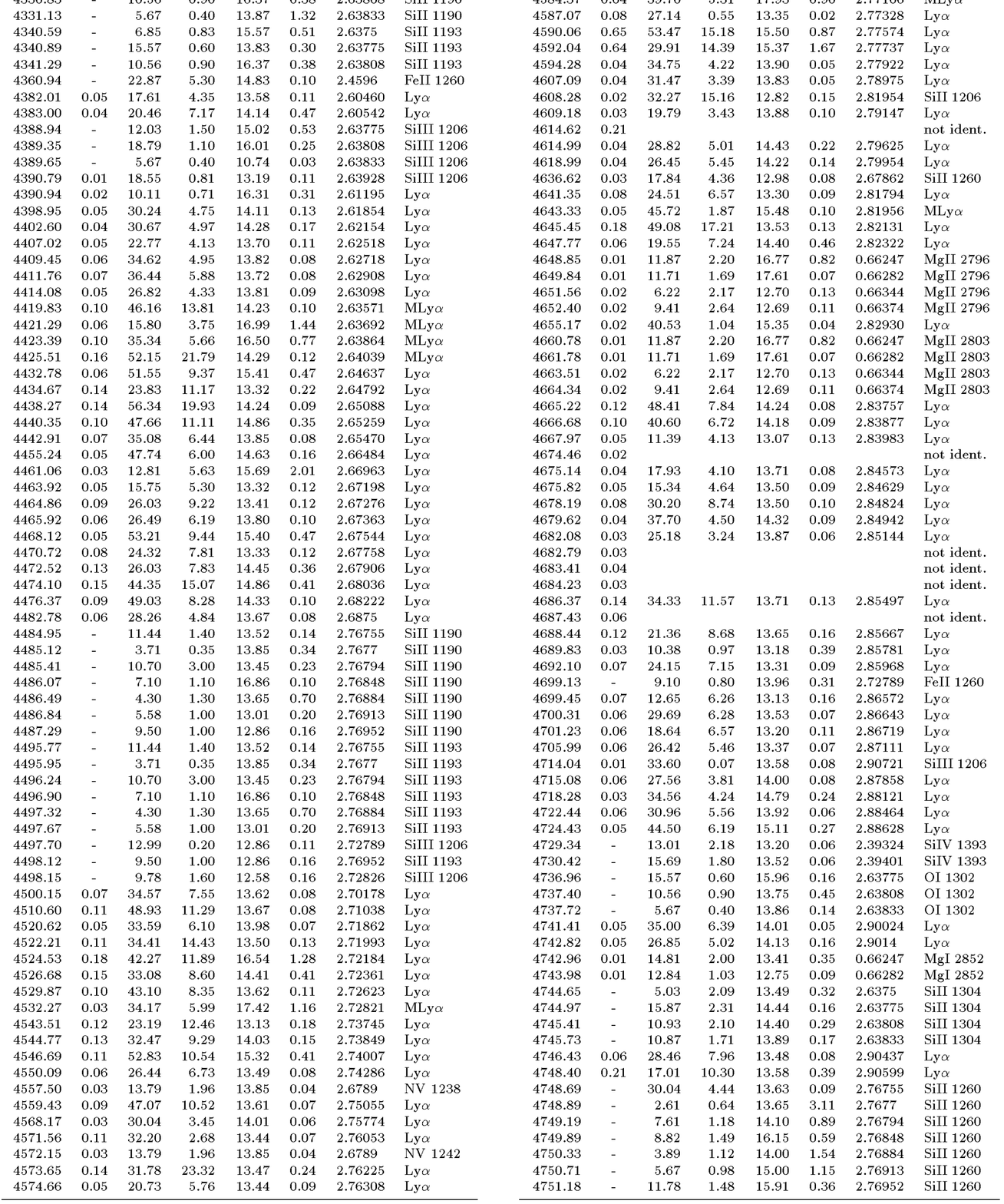}
\end{table*}

\setcounter{table}{1}
\begin{table*}\caption[t2a]{\it -- Continued}
{\label{t2a}}
\epsfysize=26cm
\epsffile{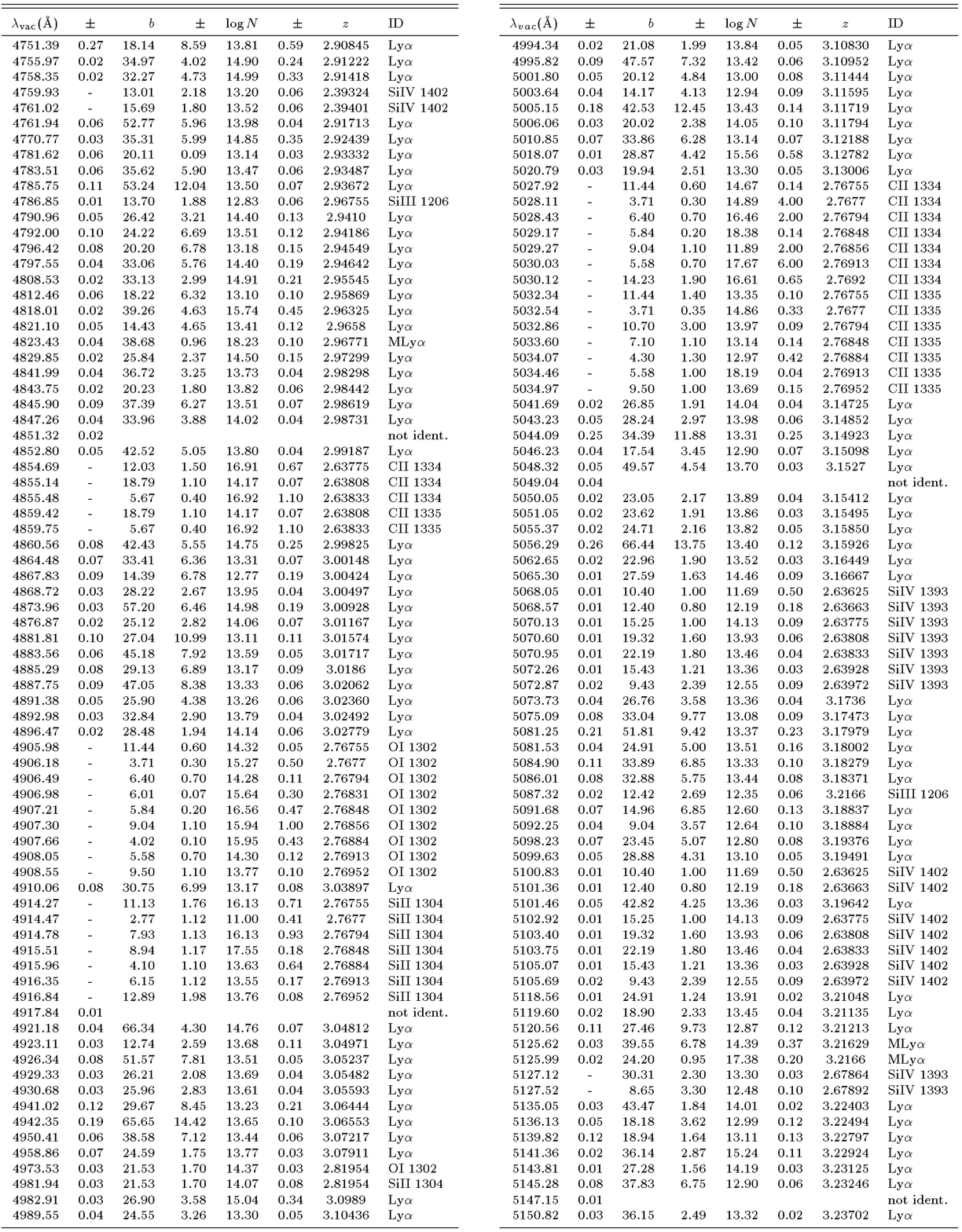}
\end{table*}

\setcounter{table}{1}
\begin{table*}\caption[t2b]{\it -- Continued}
{\label{t2b}}
\epsfysize=26cm
\epsffile{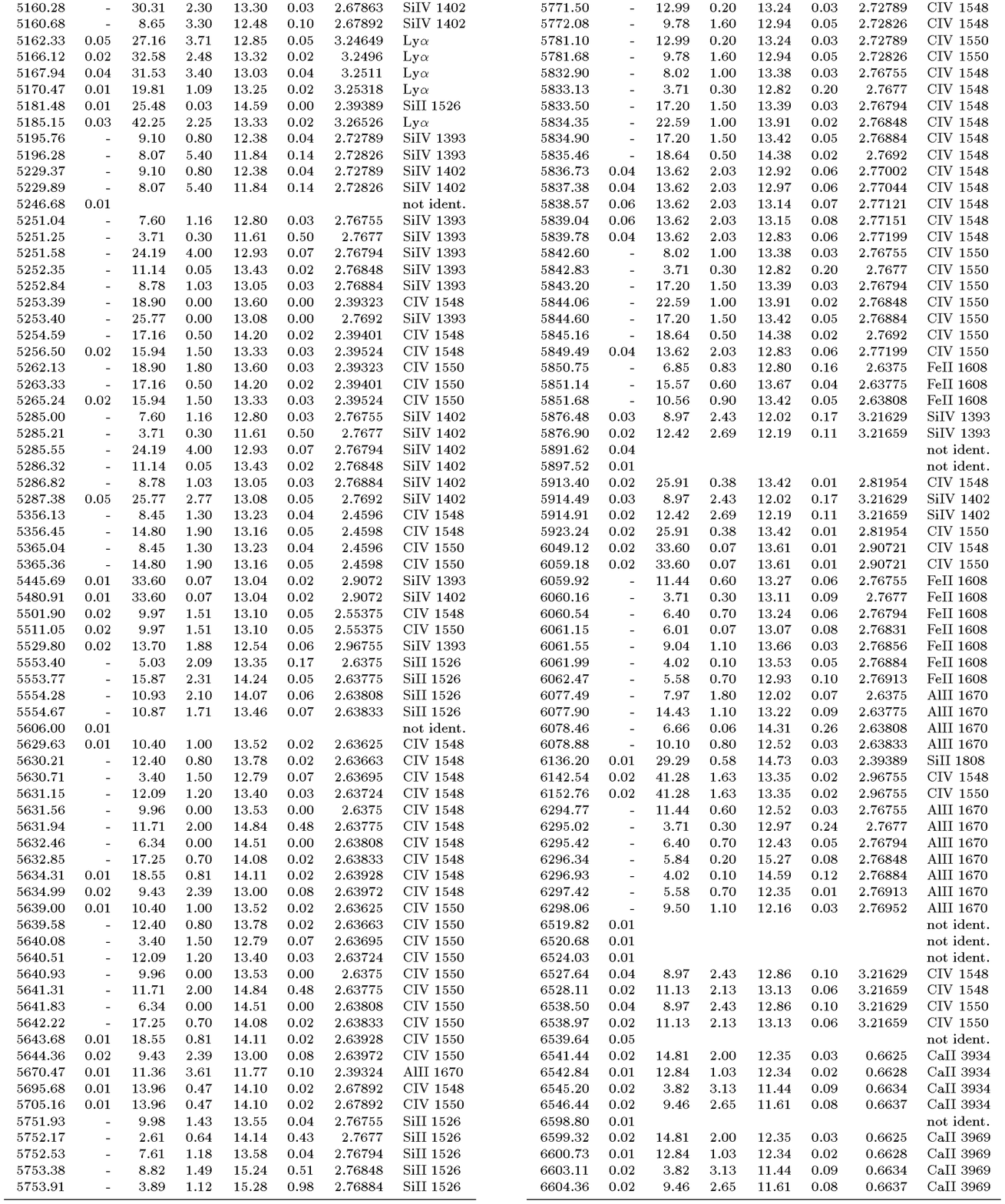}
\end{table*}

\section{Lyman alpha statistics}

A statistically well-defined sample of \Lya\ lines in the region between the
Ly$\beta$ and Ly$\alpha$ emissions ($\lambda\lambda4380-5191$) can be 
obtained from Tab.~\ref{t2}.
The \Lya\ lines affected by the proximity effect ($8 h_{100}^{-1}$ Mpc from 
the QSO; see Bajtlik et al. 1988; Lu \eal 1991) 
and those associated with metal systems (indicated as M\Lya) 
are excluded from the sample. 

The distribution of lines in the $b-\log N$ plane is shown in 
Fig.~\ref{fig:b-N}. 
The lack of lines at the top left corner 
can be ascribed 
to an observational bias: as already shown by GCFT, the selection criterion 
tends to miss lines with low column density and large Doppler width.  

The dataset can be considered virtually complete (for typical $b$ 
values) for $\log N_{\rm HI} \gsim 13.3$. 

A considerable fraction of the lines with $\log N_{\rm HI} \gsim 14.5$
belongs to complex saturated systems.
As a consequence, the deblending choices and therefore the fitting 
parameters may be not always unique. 
A considerable improvement can be obtained when     
the simultaneous fit of the saturated \Lya\ and the corresponding Ly$\beta$ 
line is possible. 
Unfortunately, due to the high density of \Lya\ lines at these redshifts, 
Ly$\beta$ absorptions with an uncontaminated profile are rare and the 
uncertainty for most of the lines in the right region of the diagram 
~\ref{fig:b-N} cannot be removed.

Simulations carried out by Fontana and Ballester (1995) show that, for isolated,
unsaturated lines at $s/n \gsim 8$, 
the parameters given by the fitting procedure are 
quite close to the ``true'' value, with a small and symmetric scatter 
around it. 
At column densities larger than $10^{14.5}$ cm$^{-2}$ the 
\Lya\ lines are saturated and $b$ and $N$ correlate strongly,
increasing the uncertainties.


\begin{figure}
\epsfxsize=90truemm
\epsffile{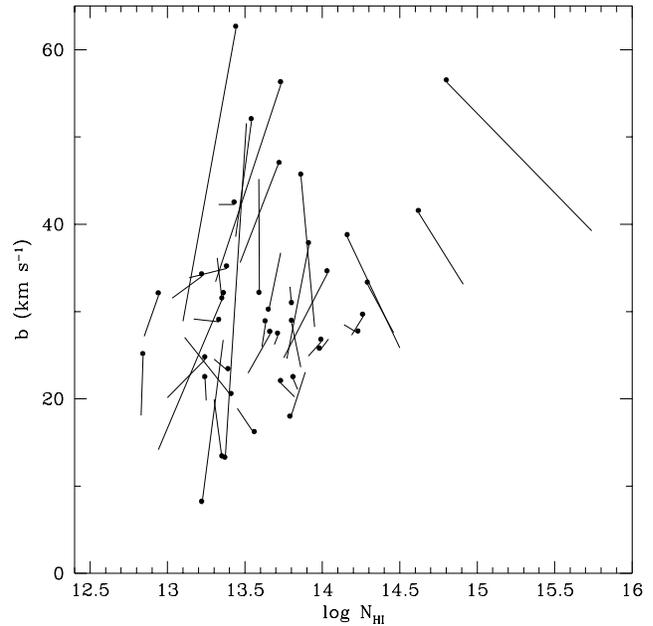}
\caption[f3]{\label{fig:migra} 
Migration diagram illustrating the effect of improving the $s/n$ 
ratio on the determination of the absorption lines parameters. The points 
correspond to the GCFT determination.} 
\end{figure}

By comparing our list of lines with that published by GCFT, we have
verified on real data how the $b-\log N$ diagram changes 
when the average $s/n$ ratio is almost doubled, checking the trends
expected on the basis of simulations.

\begin{table}
\caption[t3]{Isolated lines in common between this sample 
and that of GCFT.} {\label{t3}}
\begin{flushleft}
\begin{tabular}{rlllrrr}
\hline\hline 
&&&&&&\\[-5pt]
\multicolumn{4}{c}{Present work}&\multicolumn{3}{c}{GCFT sample} \\
\# & $\lambda_{\rm vac}$(\AA) & $\log N$ & $b$ & $\lambda_{\rm vac}$(\AA) & 
$\log N$ & $b$ \\
[2pt]\hline&&&&&&\\[-8pt]
 1 & 4783.51 & 13.47 &  35.62 &  4783.72 &   13.72 &   47.09 \\
 2 & 4808.53 & 14.91 &  33.13 &  4808.30 &   14.62 &   41.59 \\
 3 & 4818.06 & 15.74 &  39.26 &  4817.86 &   14.80 &   56.55 \\
 4 & 4829.85 & 14.50 &  25.84 &  4829.67 &   14.16 &   38.82\\ 
 5 & 4841.99 & 13.73 &  36.72 &  4841.83 &   13.65 &   30.28\\
 6 & 4843.75 & 13.82 &  20.23 &  4843.61 &   13.73 &   22.10\\
 7 & 4864.48 & 13.31 &  33.41 &  4864.25 &   13.73 &   56.34\\
 8 & 4868.72 & 13.95 &  28.22 &  4868.49 &   13.86 &   45.75\\
 9 & 4881.81 & 13.11 &  27.04 &  4881.50 &   13.41 &   20.62\\
10 & 4883.56 & 13.59 &  45.18 &  4883.55 &   13.59 &   32.21\\
11 & 4885.29 & 13.17 &  29.13 &  4885.17 &   13.33 &   29.11\\
12 & 4892.98 & 13.79 &  32.84 &  4892.90 &   13.80 &   31.03\\
13 & 4896.47 & 14.14 &  28.48 &  4896.26 &   14.23 &   27.78\\
14 & 4926.34 & 13.51 &  51.57 &  4926.28 &   13.37 &   13.32\\
15 & 4929.33 & 13.69 &  26.21 &  4929.10 &   13.71 &   27.53\\
16 & 4930.68 & 13.61 &  25.96 &  4930.46 &   13.63 &   28.94  \\    
17 & 4950.41 & 13.44 &  38.58 &  4950.44 &   13.54 &   52.10\\
18 & 4958.86 & 13.77 &  24.59 &  4958.70 &   13.91 &   37.90\\
19 & 4989.55 & 13.30 &  24.55 &  4989.60 &   13.39 &   23.47\\
20 & 4994.34 & 13.84 &  21.08 &  4994.23 &   13.81 &   22.55\\
21 & 5001.80 & 13.00 &  20.12 &  5001.91 &   13.24 &   24.82\\
22 & 5003.64 & 12.94 &  14.17 &  5003.67 &   13.36 &   32.17\\
23 & 5010.85 & 13.14 &  33.86 &  5010.40 &   13.38 &   35.23\\
24 & 5020.79 & 13.30 &  19.94 &  5020.84 &   13.35 &   13.48\\
25 & 5041.69 & 14.04 &  26.85 &  5041.48 &   13.98 &   25.82\\
26 & 5050.05 & 13.89 &  23.05 &  5049.84 &   13.79 &   18.03\\
27 & 5051.05 & 13.86 &  23.62 &  5050.83 &   13.80 &   29.00\\
28 & 5055.37 & 13.75 &  24.71 &  5055.22 &   14.03 &   34.67\\
29 & 5062.65 & 13.52 &  22.96 &  5062.51 &   13.66 &   27.73\\
30 & 5065.30 & 14.46 &  27.59 &  5065.08  & 14.29  & 33.37\\
31 & 5073.73 & 13.36 &  26.76 &  5073.34 &   13.22 &    8.26\\
32 & 5099.63 & 13.10 &  28.88 &  5099.25 &   13.44 &   62.69\\
33 & 5118.56 & 13.91 &  24.91 &  5118.41 &   13.99 &   26.83\\
34 & 5119.60 & 13.45 &  18.90 &  5119.37 &   13.56 &   16.25\\
35 & 5143.81 & 14.19 &  27.28 &  5143.64 &   14.26 &   29.70\\
36 & 5150.82 & 13.32 &  36.15 &  5150.61 &   13.35 &   31.57\\
37 & 5155.51 & 12.83 &  18.10 &  5155.42 &   12.84 &   25.19\\
38 & 5162.33 & 12.85 &  27.16 &  5162.03 &   12.94 &   32.15\\
39 & 5167.94 & 13.03 &  31.53 &  5167.80 &   13.22 &   34.33\\
40 & 5170.47 & 13.25 &  19.81 &  5170.31 &   13.24 &   22.54\\
41 & 5185.15 & 13.33 &  42.25 &  5184.94 &   13.43 &   42.56\\
[2pt]\hline
\end{tabular}
\end{flushleft}
\end{table}

To carry out a meaningful comparison, we have considered
only the isolated \Lya\ lines in common between the two line lists, 
at wavelengths $\lambda > 4750$ \AA. 
In this range the $s/n$ per resolution element is $\gsim 8$.
The lines are listed in Tab.~\ref{t3}.

Figure~\ref{fig:migra} shows the individual trajectory of each absorption 
line in the plane $b-\log N$. GCFT 
parameter values are indicated by the solid black circles and our values 
correspond to the end of the adjoining line. 

Saturated lines tend to move keeping approximately
constant the equivalent width value, as 
observed in the simulations (Fontana \& Ballester 1995). 
Often they correspond to complex features.
If the improved $s/n$ is not sufficient for a proper deblending, but
the better definition of the wings forces a fit 
with a lower $b$ parameter, then they move
toward higher column densities.

Lines with low column density in regions with low $s/n$ ratio 
have poorly defined profiles and are very sensitive to the continuum level.
The $s/n$ increase, together with a choice of a slightly lower continuum
with respect to GCFT, has induced a migration from high $b$ values and 
small column densities toward smaller Doppler widths.


\begin{figure}
\epsfxsize=90truemm
\epsffile{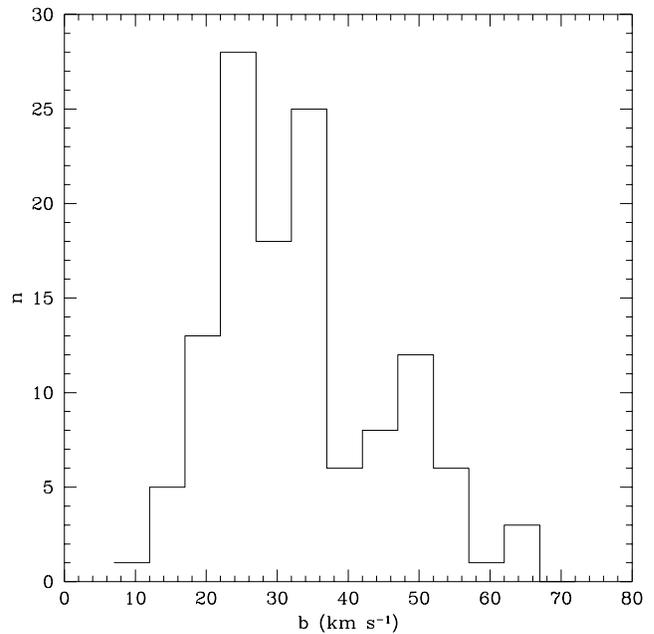}
\caption[ ]{\label{fig:bdis} 
Doppler parameter distribution for the Ly$\alpha$ lines in Tab.~\ref{t2}, 
out of 8 Mpc from the quasar PKS 2126--158.}
\end{figure}

The Doppler width distribution of the complete ($\log N_{\rm HI} \ge 13.3$) 
sample of \Lya\ lines 
is shown in Fig.~\ref{fig:bdis}.  

Using the standard assumption that line broadening is due exclusively to 
thermal motion, the relation between Doppler parameter and temperature, 
$b^2 = 2\,k\,T/m$, (where $k$ is the Boltzmann constant and $m$ is the 
hydrogen mass) implies that the peak value $b\simeq 25$ km 
s$^{-1}$ corresponds to a cloud temperature of $T\sim 38000$ K. 



The \Lya\ sample contains 21 lines, 
i.e. $\simeq14.5$ \%, with $10\leq b\leq 20$ km s$^{-1}$, and 12 lines, 
i.e. $\simeq8.3$ \%, with $15\leq b \leq 20$ km s$^{-1}$. Such percentages 
are almost halved compared with those found in the previous work by GCFT.  
Nonetheless, it has to be noted that the peak value of the $b$ 
distribution is roughly the same for the two samples.  

\vskip 12pt

The nature of the distribution of \Lya\ clouds Doppler 
parameters has led in the past to controversies.
Pettini et al. (1990) found \Lya\ lines with a median Doppler parameter of 
17 km s$^{-1}$ and a strong intrinsic correlation between Doppler width and 
column density. From these results a 
scenario of very cool, dense and practically neutral clouds emerged, 
in contrast with previous models.
Starting from data with similar resolution, Carswell et al. (1991) following 
different selection and analysis criteria obtained significantly larger 
average and median Doppler parameters and, above all, no $b-N$ correlation. 
Our median $b$ value is intermediate between the results of Pettini et al. 
and that of Carswell at al. and it agrees with recent result at very high 
resolution (Hu at al. 1995; Lu et al. 1996).

\begin{figure}
\epsfxsize=90truemm
\epsffile{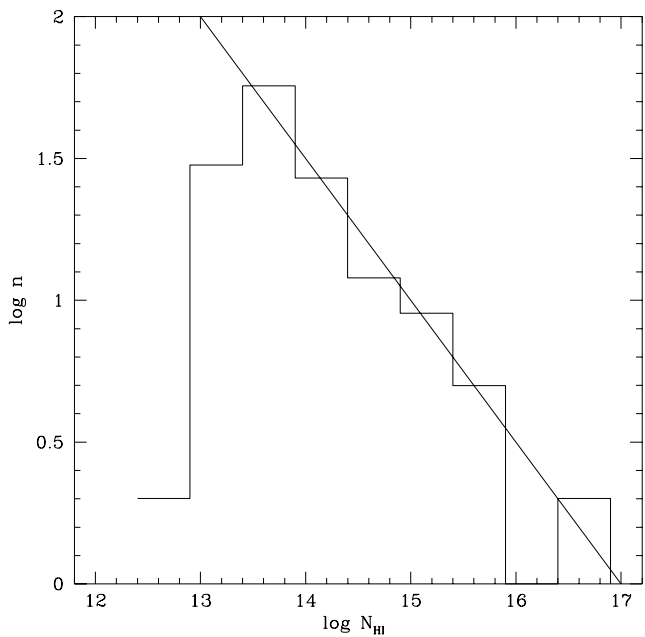}
\caption[ ]{Column density distribution of the Ly$\alpha$ lines out of 8 Mpc 
from the quasar PKS 2126--158. The overplotted solid line represents a 
power-law distribution, $\propto N^{-\beta}$, with $\beta =1.5$ (Hu et al. 
1995; Giallongo et al. 1996).}
\label{fig:NHdis} 
\end{figure}

The column density distribution of our \Lya\ line sample is shown in 
Fig.~\ref{fig:NHdis}. As shown in Fig.~\ref{fig:b-N},
for values $\log N_{\rm HI} \lsim 13.3$ a selection bias is expected: 
only lines with small Doppler parameters are detectable. This is 
confirmed by the drop of the column density distribution below this $\log 
N_{\rm HI}$ value.  
The shape of the distribution is in  
agreement with the power-law fit $N^{-\beta}$, with $\beta\simeq 1.5$, 
obtained in the recent works by Hu et al. (1995) and Giallongo et al. (1996).


The number density of lines per unit redshift, in the wavelength 
interval $\lambda\lambda 4380-5065$ \AA\ for \Lya\ absorptions 
with $\log N_{\rm HI} \ge 14$, is in good agreement 
with the cosmological redshift distribution (Giallongo \eal 1996).

\section{Identification and analysis of the metal systems}

As a first guess, heavy-element systems have been identified on the basis 
of the line list presented by GCFT. Eventually, other element lines have 
been added to the previously found systems, making use of a list of the 
lines most frequently observed in QSO absorption 
spectra, derived from Morton (1991).
The oscillator strengths of Si II $\lambda1526$ and Si II 
$\lambda1304$ have been corrected according to Verner et al. (1996).

As for the \Lya\ lines, the operation of fitting has been carried out within 
the context FITLYMAN of the reduction package MIDAS. In the case of the 
complex profiles observed for the heavy-element lines, a minimum number of 
components needed to obtain a $\chi^2 <1$ and a good fit was deduced from the 
C~IV profiles (or, in one case at low redshift, from the Mg II one). 
  
Then an identification programme, based on the method of 
Young et al. (1979), has been applied to the newly observed
lines that seemed not to belong to existing systems.  

Besides the systems previously found at lower resolutions (Young et al. 1979; 
Sargent et al. 1988; GCFT), we show two systems identified on the
basis of the C~IV doublet, whose column densities and equivalent widths were too 
weak to be detected in previous observations. We do not confirm the existence 
of the system at $z=2.33$ found by GCFT, whose lines were 
all inside the Ly$\alpha$ forest. Six of the twelve systems show a 
multicomponent substructure, with a velocity extent up to $\sim 350$ km 
s$^{-1}$. We have not adopted any velocity window, or other strict rules for 
the classification into systems versus sub-systems. We have simply
considered as one system those groups of lines for which a simultaneous fit 
was required because of the superposition of the various line profiles.

\subsection{Notes on individual systems}

Each system has been assigned a letter, from A to L, in order of increasing 
redshift. 
  
\subsubsection{The metal system at $z_{\rm abs}=0.6631$ -- A}

The two strongest Ca II doublets of this system have been already observed 
by GCFT. Furthermore, two other components have been fitted, together with the 
respective Mg II doublets. The two components with higher column density 
values show also Mg I $\lambda2852$.

The total spread in velocity of the system is $\sim 215$ km s$^{-1}$.

\subsubsection{The metal system at $z_{\rm abs}=2.3941$ -- B}

The existence of this system was formerly suggested by Young et al. and 
subsequently confirmed by Sargent et al. (1990) and by GCFT. 

The C~IV $\lambda1548$ line at $z_{\rm abs}=2.3932$ is blended with the 
Si~IV $\lambda1393$ of the complex system H, but the other line of the 
doublet has a very clean profile that guarantees the reliability of the fit.
 
We have found another component by means of the C~IV doublet, which is 
slightly separated ($\sim 100$ km s$^{-1}$) from the others. The maximum 
separation among the three is $\sim 180$ km s$^{-1}$. 

Corresponding to two lower redshift components we have fitted the Si~IV 
doublets and at $z_{\rm abs}=2.3932$ also Si~II 
$\lambda1526$, Si II $\lambda1808$ and Al~II $\lambda1670$. 

\subsubsection{The metal system at $z_{\rm abs}=2.4597$ -- C}

This system, identified by Sargent and collaborators 
(1988), shows two C~IV doublets. Relative to the stronger component, we 
observe an uncertain Fe~II $\lambda1260$ and the Ly$\alpha$, both in a 
region with low $s/n$ ratio. The Si~IV doublets, if present, are blended 
with the stronger C~II complex of system E. 

\subsubsection{The metal system at $z_{\rm abs}=2.5537$ -- D}

We have identified this low column density C~IV doublet in a region with 
high $s/n$ ratio and absence of lines. This new system does not show any 
other metallic line associated with it. 
 
\subsubsection{The metal system at $z_{\rm abs}=2.6378$ -- E}

This multicomponent system shows ten sub-features, three more 
than in GCFT, with a total spread in velocity of $\sim 286$ km s$^{-1}$. 
The number 
of components and the relative redshifts have been derived from the 
observation of the C~IV lines which lie outside the Ly$\alpha$ forest. 

Lines shortward of the Ly$\alpha$ emission are difficult to examine because 
they are blended with H~I lines. In these cases (e.g. C~II 
$\lambda1335$) we 
have identified and fitted any components in coincidence with C~IV 
components. Nevertheless, the $b$ and $N$ values of these lines, and even 
their existence, are to be considered doubtful. 

In addition to the elements found by GCFT, we have fitted Si~III 
$\lambda1206$, Si~II $\lambda1304$ and $\lambda\lambda1190,1193$, all in a 
region of the spectrum with low $s/n$ ratio. 

The corresponding Ly$\alpha$ line has been fitted separately. As a matter 
of fact, it was so strongly saturated that even the observation of the 
Ly$\beta$ did not help in the determination of its profile.  

\subsubsection{The metal system at $z_{\rm abs}=2.6788$ -- F}

A clear C~IV doublet, first identified by Sargent et al. (1988), is 
observed 
outside the Ly$\alpha$ forest; all the other lines of this system lie inside 
it. At the same redshift of C~IV, Si~IV doublet and N~V doublet 
have been found.

Si~IV doublet has been fitted using Si~IV $\lambda1402$ only, since 
Si~IV $\lambda1393$ is blended with a wide \Lya\ line. Besides, it   
shows a possible second component separated by $\sim 20$ km s$^{-1}$ from 
that identified by the C~IV. A Si~II $\lambda1260$ is observed at this 
last redshift, while the existence of the other lines found by GCFT is not 
confirmed. 

\subsubsection{The metal system at $z_{\rm abs}=2.7281$ -- G} 

This system has been identified by the C~IV doublet, which presents 
two components separated by $\sim 30$ km s$^{-1}$. 

Unlike in GCFT, Ly$\alpha$ and Si~IV doublet, even if extremely weak, have 
been fitted. Other lines, whose identification is uncertain, are: Si~III 
$\lambda1206$ blended with the multicomponent Si~II of system H and 
Fe~II $\lambda1260$ blended with a \Lya\ line. 

\subsubsection{The metal system at $z_{\rm abs}=2.7689$ -- H}

This is the most complex system found in our spectrum. It 
appears as a blended feature of $\Delta v\simeq 350$ km s$^{-1}$. 

We observe seven C~IV doublets and four possible very weak lines of 
C~IV $\lambda1548$. Four more components have been identified using 
low-ionization lines falling outside the Ly$\alpha$ forest, and all the lines 
unambiguously identified in the Ly$\alpha$ forest have been fitted. 

Besides the lines fitted by GCFT, we have added those of Si~II 
$\lambda1304$, $\lambda1260$, $\lambda1190$, $\lambda1193$ and the 
fine-structure excited level C~II $\lambda1335$. 
No fine-structure line of Si~II has been observed. 

As for the other complex system, E, the Ly$\alpha$ line has been observed but 
not fitted together with the other elements. The presence of the 
corresponding Ly$\beta$ line does not help much because of the strong 
saturation. 

\begin{figure}
\epsfxsize=90truemm
\epsffile{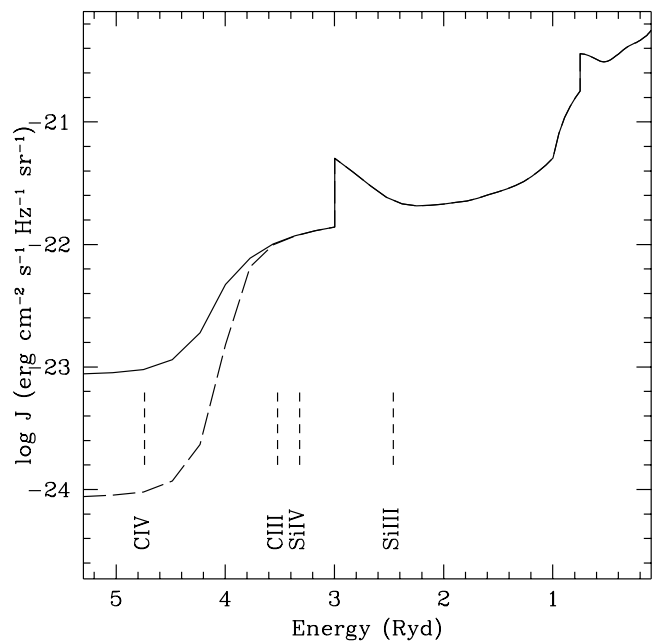}
\caption[ ]{The shape of the ultraviolet background at the epoch
z=2.9. Continuous line: as computed by Haardt and Madau (1997); dashed
line: with an increased (of a factor 10) jump around 4 Ryd. The
ionization potentials of a few ions are shown.} 
\label{fig:uvbkg}
\end{figure}

\subsubsection{The metal system at $z_{\rm abs}=2.8195$ -- I}

Meyer and York (1987) identified this system as a single C~IV 
doublet. We have added the identifications of O I $\lambda1302$, Si~III 
$\lambda1206$, Ly$\alpha$, Ly$\beta$ and Ly$\gamma$, while the existence 
of Si~II $\lambda1304$, $\lambda1309$ is uncertain due to severe blending. 

None of the low-ionization lines reported by GCFT have been observed, apart 
from Al~II $\lambda1670$ that has become a C~IV at $z=3.216542$. 

\subsubsection{The metal system at $z_{\rm abs}=2.9072$ -- J}

This C IV system (Meyer \& York 1987) shows also Si~III 
$\lambda1206$, Si~IV $\lambda\lambda1393,1402$, Ly$\alpha$ Ly$\gamma$ 
and Ly$\delta$. 

\subsubsection{The Lyman limit system at $z_{\rm abs}=2.9675$ -- K}

This is a Lyman-limit system identified by Sargent et al. (1990). 

The Ly$\alpha$ line of this system probably has a complex structure which 
has not been possible to disentangle even examining the corresponding 
Ly$\gamma$, still badly saturated. 

At high resolution, the C~IV identification is doubtful (a wavelength 
discrepancy of about $0.5$ \AA\ is observed between the two components of 
the doublet).  We have identified and fitted a very weak Si~IV 
$\lambda1393$ and Si~III (possibly contaminated by Ly$\alpha$).  
The existence of the N~V doublet is not verifiable since a wide 
H~I line is present at that wavelength. 

\subsubsection{The metal system at $z_{\rm abs}=3.216542$ -- L}

This possible system, previously unknown, shows a C~IV 
doublet (the C~IV $\lambda1548$ line was identified as Al~II $\lambda1670$ 
at $z=2.81959$ by GCFT) and a Si~IV doublet outside the Lyman forest. 
We have fitted also the Ly$\alpha$ line and Si~III $\lambda1206$ in 
the forest. 

For the systems I and J the reliability of the fit of the Lyman series
allows a measure of the metallicity. In system K it is possible 
to put an upper limit to it on the assumption of $\log {\rm N_{HI}}
\geq 17$.
We have used the standard photoionization code CLOUDY (\cite{CLOUDY}),
following the same approach described in detail in Savaglio et al.
(1997).

In the analysis of system J, if the standard UV background at
z=2.9 (\cite{HM:96}) is adopted and the size of the cloud is
assumed not smaller than 20 Kpc, we end up with an unplausibly high
overdensity of silicon with respect to carbon, $\ga 50$.
To obtain a more realistic ${\rm [Si/C]} \la 1$ the cloud size 
should decrease substantially below 10 kpc.
Incongruities of this type are not uncommon, 
as reported by Songaila and Cowie (1996) and Savaglio et al. (1997). 
Several processes may increase the
SiIV/CIV ratio: a photoionizing background dominated by local sources,
non-equilibrium temperatures and non-uniform radiation fields
(\cite{gir:shu:97}). In Savaglio et al. (1997) and here we have
explored the possibility of an enhanced break at 4 Ryd in the
metagalactic background radiation, as could be originated by 
a contribution of primeval galaxies in addition to the standard QSO background.
With a jump increased of a factor 10 at 4 Ryd (Fig.~\ref{fig:uvbkg}), 
the observed line ratios turn out to be compatible, assuming a cloud
size of 30 kpc, with a metallicity -2.5 dex solar 
and an overabundance of silicon with respect to carbon of a factor 5-6.

The analysis of systems I and K, under the assumption of the same
UV background with an enhanced jump at 4 Ryd and cloud sizes $\ga
15$ kpc, provides metallicities of $\simeq -2.2$ and $\la -2.7$,
respectively. System I, again, requires an overabundance of silicon with
respect to carbon of a factor 4 that would become much larger 
if the standard UV background is assumed.

\begin{figure}
\epsfxsize=90truemm
\epsffile{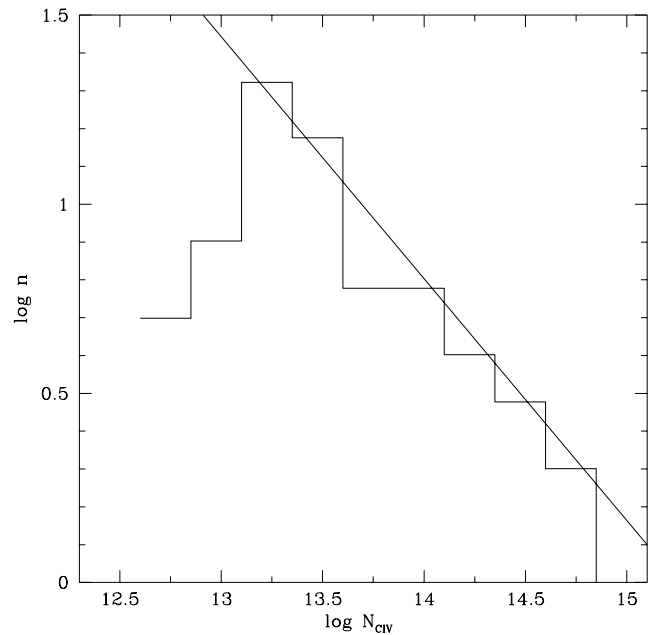}
\caption[ ]{Histogram of the logarithmic number of C~IV absorptions as a 
function of the column density. The solid line represents 
a power-law: $N_{\rm CIV}^{-\beta}$, $\beta=1.64$ (\cite{pb94}).} 
\label{fig:nchi}
\end{figure}

\section{Statistics of C~IV systems}

In order to investigate the statistical properties of 
C~IV systems, an enlarged sample has been created 
by merging the present data with those recently obtained at 
similar resolution for Q0000--26 (\cite{sava:97})
and Q0055 - 269 (Cristiani et al. 1995).
All the spectra have been analyzed in a homogeneous way according to the same
procedures described in sections 2 and 3.

\subsection{Doppler parameter and column density}

The distribution of C~IV column densities is 
shown in Fig.~\ref{fig:nchi}. 
It can be described as a single power-law distribution, $dn/dN \propto
N^{-\beta}$, with $\beta= 1.64$ (in agreement with \cite{pb94}), down to a 
completeness limit $\log N_{\rm CIV} \sim 13.3$.

\begin{figure}
\epsfxsize=90truemm
\epsffile{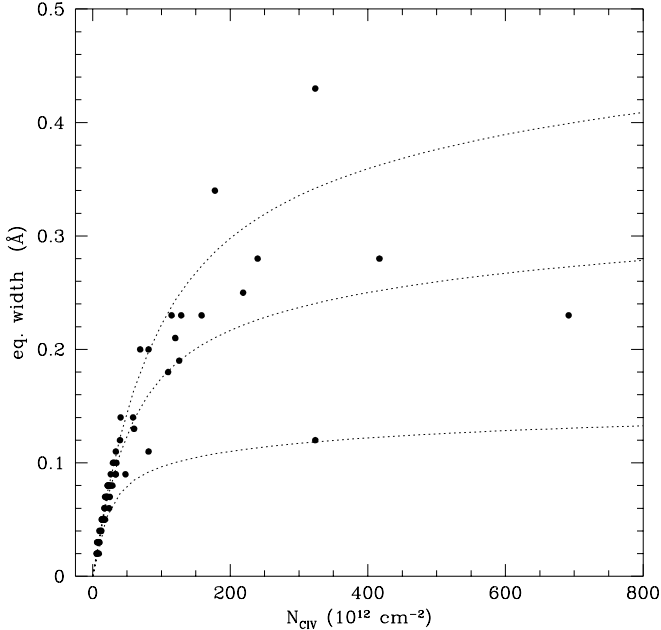}
\caption[ ]{Rest equivalent width of single C~IV system components as 
a function of the column density (in units of $10^{12}$ cm$^{-2}$). The 
overplotted curves of growth have been computed for values $b = 6, 14, 22$ km 
s$^{-1}$ of the Doppler parameter (Spitzer 1978; Press \& Rybicki 1993).} 
\label{fig:cgro} 
\end{figure}

The distribution of the Doppler parameters for the same C~IV lines has a 
mode of $\simeq 14$ km s$^{-1}$ and a standard deviation $\sigma_b 
\simeq 8$ km s$^{-1}$.
A simple derivation of the expected  $ < b_{\rm CIV} >$ from the
corresponding $ < b_{{\rm Ly}\alpha} >$, with the hypothesis of pure
thermal broadening with a single temperature would provide 
$< b_{\rm CIV} > \simeq 7 $ km s$^{-1}$.
This indicates either that turbulent motions are present or
some of the lines are blends of subcomponents, which would
be separated only by observations with higher spectral resolution
and $s/n$.

Figure~\ref{fig:cgro} shows the equivalent widths, estimated from the $b$
and $N_{\rm CIV}$ values derived for each subcomponent by profile fitting,
versus the column density. It can be seen that most of the lines are
on the linear part of the curve of growth, at least for $N_{\rm CIV} <
10^{14}$ cm$^{-2}$. 
Curves of growth have been drawn for values of the Doppler parameter 
$b_{\rm CIV} = 6,14,22$ km s$^{-1}$.

\section{Clustering of C~IV clouds}

\subsection{Statistical tools}

We have adopted
as a statistical tool the two-point correlation function (TPCF), 
defined as the excess, due to clustering, of the probability $dP$ of 
finding a cloud in a volume $dV$ at a distance $r$ from another 
cloud:

\begin{equation}
dP = \Phi(z)\,[1+\xi (r)]\,dV
\end{equation}

\noindent
where $\Phi(z)$ is the average space density of the clouds as a function 
of $z$. The TPCF is known to be a satisfactory 
estimator when used to investigate weak clustering on scales considerably 
smaller than the total interval covered by the data. The binning,  
intrinsic to this method, causes a loss of information, but the ease in 
visualizing its results and in 
including observational effects in the computing codes have made of the 
TPCF one of the favorite statistical estimators in cosmology.

In practice the observations provide the redshifts of the absorption 
lines that, due to peculiar motions, are not immediately transformed in 
comoving distances. Therefore the TPCF is generally computed in the 
velocity space, making use of the formula (Peebles 1980)

\begin{equation}
\xi (v, \Delta v) = \frac{n_{\rm obs}(v, \Delta v)}{n_{\rm exp}
(v, \Delta v)}-1 
\end{equation}

\noindent
where $n_{\rm obs}$ is the number of observed line pairs with velocity 
separations between $v$ and $v+\Delta v$ and $n_{\rm exp}$ is the number 
of pairs expected in the same interval from a random distribution in 
redshift.

At the small velocity separation we are dealing with, the variation of 
the distance scale with cosmic time can be neglected and the velocity 
difference can be simply deduced from the redshift difference (Sargent et 
al. 1980)

\begin{equation}
\Delta v = \frac{c\,(z_2 -z_1)}{1+(z_1+z_2)/2}
\end{equation}

\noindent
where $\Delta v$ is the velocity of one cloud as measured by an observer 
in the rest frame of the other. 

In our line sample $n_{\rm exp}$ is obtained by averaging 5000 numerical 
simulations of the observed number of redshifts, trying to account for all 
the relevant cosmological and observational effects. In particular the 
set of redshifts is randomly generated in the same redshift interval as 
the data according to the cosmological distribution $\propto 
(1+z)^{\gamma}$, where $\gamma$ has been taken equal to $-1.2$ (Sargent 
et al. 1988). 
Observed pairs with a velocity splitting $\Delta v <30$ km s$^{-1}$ have 
been merged, while simulated ones have been excluded in the estimate of 
$n_{\rm exp}$, because of the intrinsic line blending due to the typical 
widths of the C~IV lines.

\begin{figure}
\epsfxsize=90truemm
\epsffile{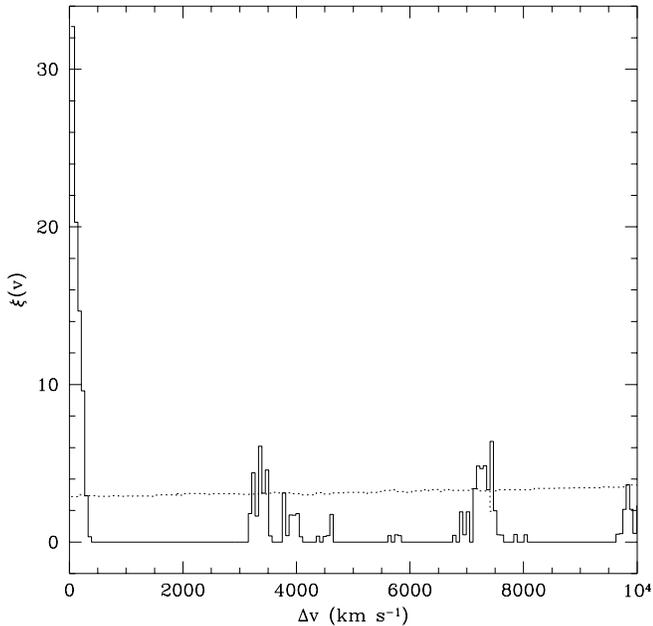}
\caption[ ]{Two point correlation function at large velocity separations  
for the C~IV systems. The dotted line shows the 95 \% confidence level.} 
\label{fig:corr1} 
\end{figure}

\begin{figure}
\epsfxsize=90truemm
\epsffile{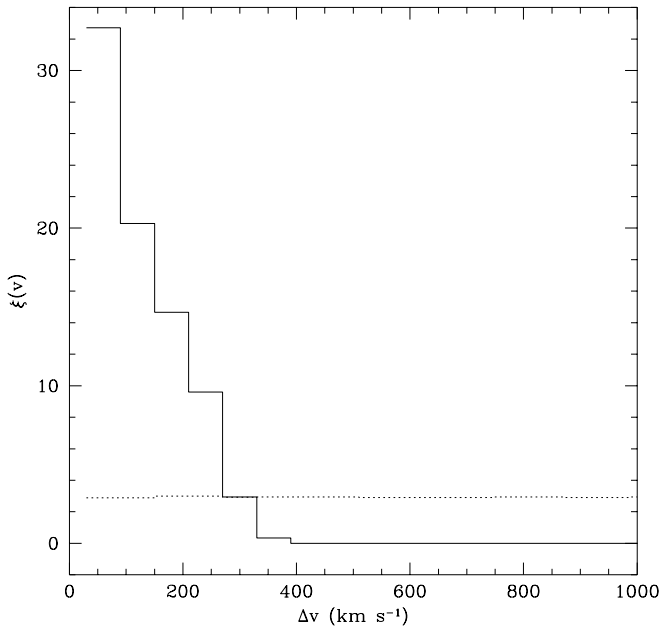}
\caption[ ]{Two point correlation function at small velocity separations 
for the C~IV systems. The dotted line shows the 95 \% confidence level}
\label{fig:corr2}
\end{figure}

The TPCF for the C~IV clouds is presented in 
Fig.~\ref{fig:corr1} with 60 km s$^{-1}$ bins. A strong clustering signal 
is detected at small velocity separations ($\Delta v < 1000$ km s$^{-1}$).
At larger scales no significant signal is found, in particular the peaks 
observed in the TPCF at $\sim 3360$ km s$^{-1}$, $\sim 7140$ km s$^{-1}$ and 
at $\sim 9840$ km s$^{-1}$ are 
aliases corresponding to the coupling of the low $\Delta v$ (high 
frequency) power with the window function.

The present data have adequate resolution to allow a further 
investigation of the distribution on scales smaller than 1000 km s$^{-1}$. 
The TPCF for velocity separations in the range $30 \le \Delta v \le 1000$ is 
shown in Fig.~\ref{fig:corr2}. In this velocity interval 48 pairs are
observed, while $\sim 10$ are predicted for a homogeneous distribution. 

Similar results, with a significant correlation on scales up to 200-300 
km s$^{-1}$, have been obtained in the works by Petitjean and Bergeron 
(1994), Womble et al. (1996) and Songaila and Cowie (1996), carried out at 
comparable resolution. 

The velocity scale at which the maximum clustering signal is observed is 
comparable with the extension of the complex metal absorption features 
in the spectrum. This suggests, as already noticed by Petitjean and Bergeron 
(1994), that we are not seeing clustering of ``galaxies'' but 
of gas clouds within the same galactic halo.

Other authors (Sargent et al. 1988; Heisler et al. 1989) report, on the 
basis of large samples of QSOs observed at relatively low resolution (a 
few Angstrom), significant correlation of C~IV lines up to scales of 
600 - 1000 km s$^{-1}$. If the resolution of the present data is degraded 
to the level typical of those investigations (e.g.a resolution of 77
km s$^{-1}$), a compatible result is 
obtained, with $\xi (0< \Delta v<1000\ {\rm km\ s}^{-1})\simeq 2.0 \pm 0.9$. 

The correlation on scales larger than 1000 km s$^{-1}$ reported by Heisler 
et al. (1989) is not reproduced. It has to be noted, however, that
it appears to be the result of the inclusion in their sample of 
one ``exceptional'' object, 0237-233. 

\subsection{Correlations with the column density}

\begin{figure}
\epsfxsize=90truemm
\epsffile{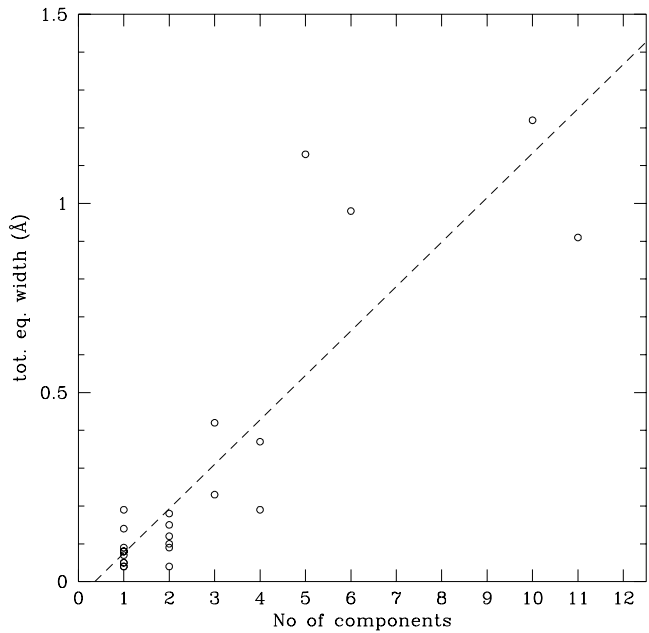}
\caption[ ]{Total equivalent width of C~IV systems  
versus the corresponding number of components. The dashed line is the 
linear correlation best fit for all points.} \label{fig:ew1}
\end{figure}

\begin{figure}
\epsfxsize=90truemm
\epsffile{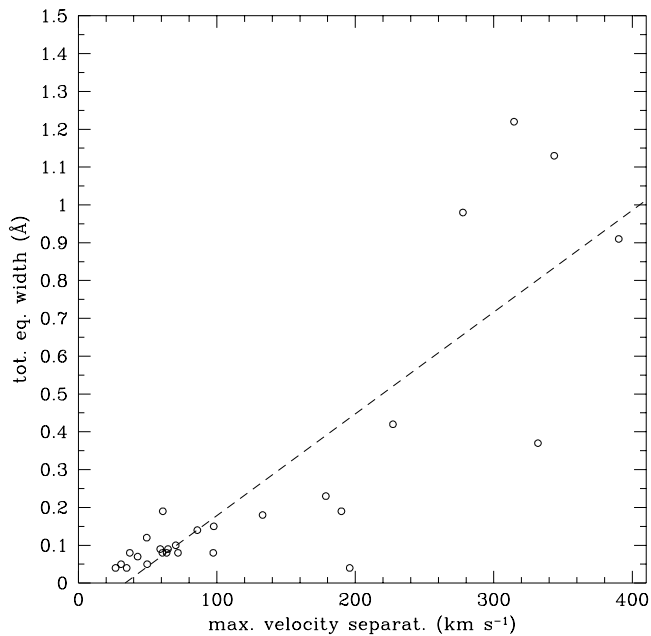}
\caption[ ]{Total equivalent width of C~IV systems  
versus the corresponding maximum velocity separation. The dashed line is 
the linear correlation best fit.} \label{fig:ew2}
\end{figure}

\begin{figure}
\epsfxsize=90truemm
\epsffile{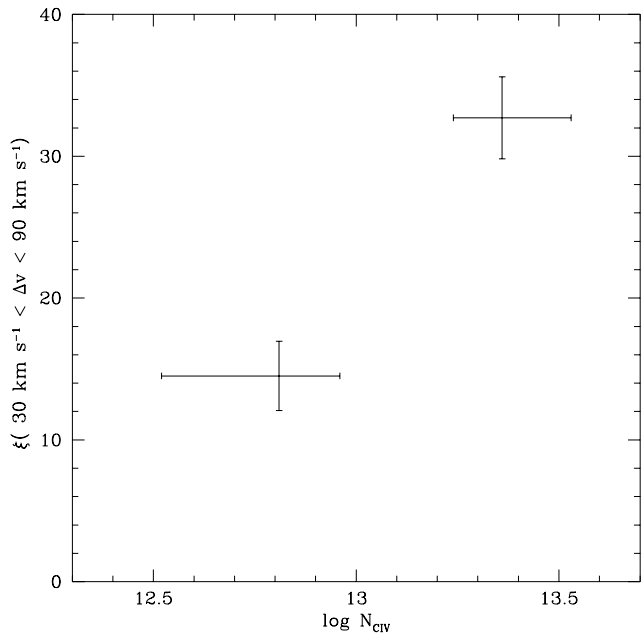}
\caption[ ]{Amplitude of the clustering of C~IV lines as a function of
the column density. The upper right point has been obtained for the 
present enlarged sample of C~IV lines.
The lower left point has been derived from the data of the QSO 1422+231
(Songaila \& Cowie 1996).
The vertical bars represent $1 \sigma$ poissonian uncertainties in the 
determination of the TPCF while the horizontal ones show the
$68 \%$ confidence interval of the column densities.} 
\label{fig:corrcd} 
\end{figure}

It has long been assessed that a correlation exists between the equivalent 
width of metal absorption systems and their number of components 
(Wolfe 1986; Petitjean \& Bergeron 1990; \cite{pb94}).

In Fig.~\ref{fig:ew1} the total equivalent width  is 
plotted versus the number of components for the C~IV absorption systems.
The dashed line represents the best linear fit for all the points. 

The number of components, however, is highly dependent on the spectral 
resolution and on subjective taste. 
To overcome this problem, we examined 
the maximum velocity separation in a given system versus the total
equivalent width (Fig.~\ref{fig:ew2}). A correlation is apparent and
the dashed line is, again, the best linear fit.

The observed trends suggest that the clustering amplitude of the C~IV lines
could be a function of the column density.

In order to further investigate this issue, in 
Fig.~\ref{fig:corrcd} the amplitude of the TPCF 
(in the bin $30 < \Delta v < 90 $ km s$^{-1}$) is plotted as a 
function of the median value of the C~IV column density.
The upper right point represents the result obtained in this work 
for the C~IV absorptions in the enlarged sample. 
The lower left point comes from the lower column density sample 
obtained by Songaila and Cowie (1996) for the QSO 1422+231. The 
TPCF has been computed, also for the latter data, according to the procedure 
described in \S~7.1.
Lines of lower column density show indeed a smaller amplitude
of the TPCF at $ 30 < \Delta v < 90 $ km s$^{-1}$.

The correlation of the clustering amplitude
with column density is analogous to what has been observed for \Lya\ 
lines (Cristiani et al. 1997) and it is qualitatively
consistent with a picture of gravitationally induced correlations.

\section{Conclusions}

We have presented high-resolution spectra of 
the quasar PKS 2126--158.

The analysis of the \Lya\ forest 
shows that:

\begin{enumerate}
\item The peak value of the Doppler parameter distribution is $\sim 25$ km 
s$^{-1}$. 
This value is large enough to be in agreement 
with a standard ionization model, but not too high 
($b \gsim 30$) 
to make the \Lya\ clouds contribution to baryonic matter 
exceed the standard nucleosynthesis value (Hu et al. 1995).

\item The plot of the statistical sample of \Lya\ lines in the $b-\log N$
plane shows no significant correlation between the two parameters. 

\item The column density distribution shows a cutoff, due to
incompleteness and blending, for $\log N_{\rm HI} \lsim 13.3$, 
for higher values it is well described by a power-law
$dn/dN_{\rm HI} \propto N_{\rm HI}^{-\beta}$ with $\beta=1.5$ 
(Hu et al. 1995; Giallongo et al. 1996).

\item The number density of lines per unit redshift 
is in agreement with a standard power-law evolution of 
the type $dn / dz \propto (1+z)^{\gamma}$ 
with $\gamma=2.7$ (Giallongo et al. 1996).
\end{enumerate}

Clustering properties of the \Lya\ lines have been investigated in another 
paper (Cristiani et al. 1997). \Lya\ of higher column density 
($\log N_{\rm HI}\gsim 13.8$) are found to cluster significantly 
on scales around 100 km s$^{-1}$. 

\vskip 12pt

From the analysis of metal absorptions the following results have been
obtained: 

\begin{enumerate}
\item Two new C~IV absorption systems have been detected at $z=3.2165$
and $z=2.5537$.
\item A mean metallicity of $\sim -2.5$ dex solar has been found 
using the metal systems at $z=2.8195$, $z=2.9072$ and 
$z=2.9675$. 
In order to make the column densities of the intervening
systems compatible 
with realistic assumptions about the cloud sizes and the [Si/C] ratios,
it is necessary to assume 
an increase of a factor 10 in the jump at 4 Ryd of the standard spectrum 
of the ionizing UV background (Haardt \& Madau 1997). 
\item Merging the present data with those obtained 
at comparable resolution for the two 
quasars Q0055-26 and Q0000-26, has provided a relative large sample
(71 doublets) of C~IV absorptions, complete down to a
column density $\log N_{\rm CIV} \simeq 13.3$.
The C~IV column density distribution is well fitted by a 
single power-law, with $\beta=1.64$.
\item The mode of the Doppler parameter distribution is $b_{\rm CIV} \simeq
14$ km $^{-1}$. 
\item The clustering properties of the individual components of the C~IV  
features have been investigated making use of a TPCF in the velocity space.  
A significant signal is obtained for scales smaller than $200-300$ km 
s$^{-1}$, $\xi(30< \Delta v < 90 {\rm km\ s}^{-1}) = 32.71 \pm 2.89$.
The result is consistent with previous findings 
(Sargent et al. 1988; Heisler et al. 1989) when the
increased resolution and $s/n$ ratio of the present data is taken into account. 
\item Correlations between total equivalent widths and number of
components and between total equivalent widths and velocity spread of
the individual C~IV systems are observed.
The two-point correlation functions for lower column density C~IV 
absorption systems, recently estimated by Womble et al. (1997) and
Songaila and Cowie (1996), show a weaker signal than the present data.
These three pieces of evidence suggest a trend of decreasing 
clustering amplitude with decreasing column density.
An analogous behaviour has been observed 
for \Lya\ lines by Cristiani et al. (1997).
\end{enumerate}

\begin{acknowledgements}
We thank P. Madau, C. Porciani and S. Savaglio for enlightening discussions.
B
SC acknowledges the support of the ASI contract 95-RS-38 and 
of the TMR-network of the European Community ``Galaxy Formation and 
Evolution''.
 
\end{acknowledgements}

\bigskip\bigskip\bigskip
A

\begin{thebibliography}{}

   \bibitem[]{}
        Bajtlik S., Duncan R.C., Ostriker J.P., 1988, ApJ 327, 570

   \bibitem[]{}
        Bechtold J., Green R.F., York D.G., 1987, ApJ 312, 50

   \bibitem[]{}
	Carswell R.F., Whelan J.A.J., Smith M.G., Boksenberg A. and
        Tytler D. 1982, MNRAS 198, 91 

   \bibitem[]{}
        Carswell R.F., Lanzetta K.M., Parnell H.C., Webb J.K., 1991, ApJ 371,
        36

   \bibitem[]{}
        Chernomordik V.V., 1995, ApJ 440, 431

   \bibitem[]{}
        Cowie L.L., Songaila A., Kim T., Hu E.M., 1995, AJ 109, 1522

   \bibitem[]{}
        Cristiani S., D'Odorico S., Fontana A., Giallongo E., Savaglio S., 
        1995, MNRAS 273, 1016

   \bibitem[Cristiani et al. 1996]{cluster96}
        Cristiani S., D'Odorico S., D'Odorico V., et al., 
	1997, MNRAS 285, 209


   \bibitem[D'Odorico 1990]{dodo90} 
	D'Odorico S., 1990, ESO The Messenger 61, 51

   \bibitem[Ferland 1996]{CLOUDY} 
	Ferland G.J., 1996, Hazy, a Brief Introduction to
	Cloudy, University of Kentucky Department of Physics and
	Astronomy Internal Report

   \bibitem[Fern\'andez-Soto et al. 1996]{fsoto}
	Fern\'andez-Soto A., Lanzetta K.M., Barcons X., et al., 
	1996 ApJ 460, L85
 
   \bibitem[Fontana \& Ballester 1995]{fitly95}
	Fontana A., Ballester P., 1995, ESO The Messenger 80, 37

   \bibitem[1993]{GCFT93} 
        Giallongo E., Cristiani S., Fontana A., Trevese D., 1993, ApJ
        416, 137

   \bibitem[]{}
        Giallongo E., Cristiani S., D'Odorico S., Fontana A., Savaglio S., 
        1996, ApJ 466, 46

   \bibitem[Giroux and Shull 1997]{gir:shu:97}
	Giroux M.L., Shull J. M., 1997 preprint astro-ph/9701160

   \bibitem[Haardt and Madau 1996]{HM:96}
	Haardt F., Madau P., 1996, ApJ 461, 20

   \bibitem[]{}
	Heisler J., Hogan C., White S.D.M. 1989, ApJ 347, 52
 
   \bibitem[]{}
	Hu E.M., Kim T., Cowie L.L., Songaila A., Rauch M. 1995, AJ 
	110, 1526

   \bibitem[]{}
        Jauncey D.L., Wright A.E., Peterson B.A., Condon J.J., 1978, ApJ 223, 
        L1
 

   \bibitem[1995]{Lanz95}
        Lanzetta K.M., Bowen D.V., Tytler D., Webb J.K., 1995, ApJ 442, 538

   \bibitem[]{}
        Lu L., Wolfe A.M., Turnshek D.A., 1991, ApJ 367, 19

   \bibitem[]{}
	Lu L., Sargent W.L.W., Womble D.S., Takada-Hidai M. 1996, ApJ  
        472, L509

   \bibitem[]{}
        Lynds C.R., 1971, ApJ 174, L73

   \bibitem[]{}
        Meiksin A., Bouchet R.F., 1995, ApJ 448, L85

   \bibitem[]{}
        Meyer D.M., York D.G., 1987, ApJ 315, L5

   \bibitem[]{}
        Morton D.C., 1991, ApJS 77, 119

   \bibitem[]{}
	Peebles P.J.E., 1980, The Large Scale Structure of the Universe.
	Princeton Univ. Press, Princeton

   \bibitem[]{}
        Petitjean P., Bergeron J., 1990, A\&A 231, 309

   \bibitem[Petitjean \& Bergeron 1994]{pb94}
        Petitjean P., Bergeron J., 1994, A\&A 283, 759


   \bibitem[]{}
        Pettini M., Hunstead R.W., Smith L.J., Mar D.P., 1990, MNRAS 246, 545
   
   \bibitem[]{}
        Press W.H., Rybicki G.B., 1993, ApJ 418, 585


   \bibitem[]{}
        Rauch M., Carswell R.F., Webb J.K., Weymann R.J., 1993, MNRAS 260, 589

   \bibitem[]{} 
        Sargent W.L.W., Young P.J., Boksenberg A., Tytler D., 1980, ApJS 42, 
        41

   \bibitem[]{} 
        Sargent W.L.W., Young P.J., Schneider D.P., 1982, ApJ 256, 374
   
   \bibitem[]{}
        Sargent W.L.W., Boksenberg A., 1983. In: 24th Li\`ege Astrophysical 
        Colloquium. 518, Quasar and Gravitational Lenses

   \bibitem[]{}
        Sargent W.L.W., Boksenberg A., Steidel C.C., 1988, ApJS 68, 539

   \bibitem[]{}
        Sargent W.L.W., Steidel C.C., Boksenberg A., 1989, ApJS 69, 703

   \bibitem[]{}
        Sargent W.L.W., Steidel C.C., Boksenberg A., 1990, ApJ 351, 364

 
   \bibitem[Savaglio et al. 1997]{sava:97}
        Savaglio S., Cristiani S., D'Odorico S., et al., 
        1997, A\&A 318, 347
   
   \bibitem[]{}
	Songaila A., Cowie L.L., 1996, AJ 112, 335

   \bibitem[]{}
	Spitzer L. Jr, 1978, Physical Processes in the Interstellar 
	Medium. John Wiley and Sons, New York, 52

 
   \bibitem[Stone \& Baldwin 1983]{sto:bal83}
	Stone R.P.S., Baldwin J.A., 1983, MNRAS 204, 347

   \bibitem[]{}
       	Tytler D., Fan X-M., Burles S., et al., 1995, in: Meylan G. (ed.) 
	QSO Absorption Lines, ESO Astrophysics Symposia, Springer: 
	Heidelberg, 289

   \bibitem[]{}
	Verner D.A, Verner E.M., Ferland G.J., 1996, BAAS 188, 5418



   \bibitem[]{}
       	Webb J.K., Barcons X., 1991, MNRAS 250, 270   

   \bibitem[]{}
	Wolfe A., 1986, in: Bregman J., Lockman J. (eds.) Proc. 
	NRAO Conf. on Gaseous Halos of Galaxies. NRAO SP, 259

   \bibitem[]{}
	Womble W.S., Sargent W.L.W., Lyons R.S., 1996, in: Bremer M.N., 
	(eds.) Proc. of the Kluwer Conf. on Cold Gas at High Redshift
   
   \bibitem[]{} 
       Young P.J., Sargent W.L.W., Boksenberg A., Carswell R.F., Whelan 
       J.A.J., 1979, ApJ 229, 891
        
   \bibitem[]{}
       Young P.J., Sargent W.L.W., Boksenberg A., 1982, ApJS 48, 455

\end{thebibliography}
\end{document}